\documentclass[a4paper,11pt]{article}
\usepackage{jheppub}
\usepackage{graphicx,placeins}
\usepackage{wrapfig}
\usepackage[utf8]{inputenc}

\newcommand\vx{{\vec{x}}}

\newcommand{\bra}{\langle}
\newcommand{\ket}{\rangle}
\newcommand{\lk}{\left(}
\newcommand{\rk}{\right)}

\newcommand{\ldk}{\left[}

\newcommand{\rdk}{\right]}
\newcommand\beq{ \begin{eqnarray} }
\newcommand\eeq{ \end{eqnarray} }

\def\lsim{\raise0.3ex\hbox{$<$\kern-0.75em\raise-1.1ex\hbox{$\sim$}}}
\def\gsim{\raise0.3ex\hbox{$>$\kern-0.75em\raise-1.1ex\hbox{$\sim$}}}

\title{\boldmath Relative scale setting for two-color QCD with Nf=2 Wilson fermions}

\author[a]{Kei Iida,}
\author[a,b,c]{Etsuko Itou} 
\author[a,d]{and Tong-Gyu Lee}%\note{}

\affiliation[a]{Department of Mathematics and Physics, Kochi University, 2-5-1 Akebono-cho, Kochi 780-8520, Japan}
\affiliation[b]{Department of Physics, and Research and Education Center for Natural Sciences, Keio University, 4-1-1 Hiyoshi, Yokohama, Kanagawa 223-8521, Japan}
\affiliation[c]{Research Center for Nuclear Physics (RCNP), Osaka University, 10-1 Mihogaoka, Ibaraki, Osaka 567-0047, Japan}
\affiliation[d]{Kochi Technical Center, e-Jan Networks Co., 4-2-52 Honmachi, Kochi 780-0870, Japan}
\emailAdd{keiiida@riken.jp}
\emailAdd{itou@yukawa.kyoto-u.ac.jp}
\emailAdd{tonggyu.lee@yukawa.kyoto-u.ac.jp}

\abstract{%
We determine the scale setting function and the pseudo-critical temperature on the lattice in $N_f=2$ two-color QCD using the Iwasaki gauge and Wilson fermion actions.
Although two-color QCD does not correspond to the real world, it is very useful as a good testing ground for three-color QCD.
The scale setting function gives the relative lattice spacings of simulations performed at different values of the bare coupling.
It is a necessary tool for taking the continuum limit.
Firstly, we measure the meson spectra for various combinations of ($\beta,\kappa$) and find a line of constant physics in $\beta$--$\kappa$ plane.
Next, we determine the scale setting function via $w_0$ scale in the gradient flow method.
Furthermore, we estimate the pseudo-critical temperature at zero chemical 
potential from the chiral susceptibility.
Combining these results, we can discuss the QCD phase diagram in which both axes are given by dimensionless quantities, namely, the temperature normalized by  the pseudo-critical temperature on the lattice and the chemical potential normalized by the pseudoscalar meson mass.
It makes it easy to compare among several lattice studies and also makes it possible to compare theoretical analyses and lattice studies in the continuum limit.
}
\begin{document} 
\maketitle
\flushbottom

%%%%%%%%%%%
%     Intro
%%%%%%%%%%%
\section{Introduction}
\label{sec:intro}
One important task in lattice quantum chrome dynamics (QCD) is scale setting.
It converts dimensionless lattice quantities to the corresponding observable in physical units.
In fact, QCD has a limited number of scales, namely, overall dynamical scale and fermion masses. 
Fixing an overall dynamical scale corresponds to finding the lattice spacing in physical units from some reference scales.
By adjusting lattice parameters so as to reproduce experimental measurements, we can obtain the scale of the lattice spacing.
On  the other hand, a related important task is to accurately determine the relative lattice spacings of simulations performed at different values of the bare coupling.
This relative scale setting makes it possible to carry out a continuum extrapolation. 
Such a relative relation between the lattice spacing and the lattice bare coupling constant is called the scale setting function.
%For instance, in the case of $N_f=2+1$ QCD,  three dimensionless reference quantities, {\it e.g.} the ratio of meson masses, calculated by the lattice QCD are used to set a scale.
%A set of lattice parameters, lattice bare coupling constant and $u,d$-quark mass and $s$-quark mass on the lattice,  is fixed to reproduce the experimental measurement, and then the scale setting function gives a prediction in physical units to other physical quantities.

The scale setting and finding the scale setting function of two-color QCD have not yet been investigated so much though two-color QCD has been well-studied as a good testing ground for three-color QCD~\cite{Nakamura:1984uz, Hands:1999md, Aloisio:2000if,Hands:2000ei, Muroya:2000qp, Kogut:2001if, Kogut:2001na, Kogut:2002cm, Muroya:2002ry, Kogut:2003ju, Hands:2006ve, Hands:2010gd, Cotter:2012mb, Boz:2013rca, Makiyama:2015uwa, Braguta:2016cpw, Holicki:2017psk, Bornyakov:2017txe, Astrakhantsev:2018uzd, Boz:2018crd, Wilhelm:2019fvp, Iida:2019rah, Boz:2019enj, Bornyakov:2020kyz, Buividovich:2020dks, Astrakhantsev:2020tdl, Takahashi:2009ef, Caselle:2015tza, Berg:2016wfw, Giudice:2017dor,Hirakida:2018uoy}. 
One of the reasons why it has not been done is that two-color QCD does not describe the real world, so that there is no reference scale in physical units.
In three-color QCD, it is possible to write the physical quantity in physical units, {\it e.g.} MeV, by comparison with the experimental data, but this procedure cannot be done in two-color QCD.
Therefore, the scale setting has often not been given much attention.
However, it is still valuable to determine the relative scale setting nonperturbatively.
It makes it doable to take the continuum extrapolation.
Furthermore, it gives a relative scale among physical observables.
Especially when there are more than two independent typical scales in the system, for instance, the critical temperature for chiral restoration
and a superfluid onset scale of the chemical potential in the finite-temperature and finite-density system, it is important to determine the relative typical scales of nonperturbative dynamics.
%Furthermore, it is useful to see the phase diagram of two-color QCD.

Actually, one attractive property of two-color QCD is that it is
free of the sign problem due to the pseudo-reality of quarks.
It is possible to perform first-principles calculation even in low-temperature and high-density regions by incorporating a diquark source term into the QCD action~\cite{Kogut:2001na}.
Several groups have independently performed first-principles calculationsin recent years~\cite{Kogut:2001na, Kogut:2002cm, Kogut:2003ju, Hands:2006ve, Hands:2010gd, Cotter:2012mb, Boz:2013rca, Makiyama:2015uwa, Braguta:2016cpw, Holicki:2017psk, Bornyakov:2017txe, Astrakhantsev:2018uzd, Boz:2018crd, Wilhelm:2019fvp, Iida:2019rah, Boz:2019enj, Bornyakov:2020kyz, Buividovich:2020dks, Astrakhantsev:2020tdl}.
Moreover, the construction of phase diagram by using a new anomaly matching for massless two-color QCD~\cite{Furusawa:2020qdz} and by using several effective theories ~\cite{Kogut:1999iv, Adhikari:2018kzh, Contant:2019lwf} has also been made.
These lattice and theoretical studies have elucidated the two-color QCD phase diagram in the finite density region, and have found rich phase structures in a manner depending on temperature and chemical potential. 
For example, even in the low-temperature region, which is considered to be lower than the critical temperature at zero chemical potential ($T_c$), there appear a quark-gluon plasma (QGP) phase, a BCS phase with color deconfinement property and a BCS phase with color confinement property with decreasing temperature at a sufficiently high chemical potential.
On the other hand,  it is difficult to make a quantitative comparison among these works because the continuous limit has not yet been taken in the lattice studies.
Moreover, it is also difficult to compare the results even among several lattice studies alone since the lattice action is different from each other.

In these situations,  the two-color QCD phase diagram on {\it normalized temperature} and {\it normalized chemical potential} plane is more useful than the one drawn in physical units.
There are two typical scales related to each axis in the two-color QCD phase diagram.
The first one is the onset scale of the quark chemical potential, which is the starting point at which the quark number density becomes non-zero at zero temperature.
It is almost the same as a half of the pseudoscalar (PS) meson mass ($m_\mathrm{PS}$), which is the lightest hadron mass at zero chemical potential.
The second one is the critical temperature at zero chemical potential ($T_c$).
Both scales in physical units depend on the number and mass of fermions of the system.
Even more generally, to compare phase diagrams between different colors and/or fermion masses, it would be useful to describe the normalized parameters,
namely, the quark chemical potential $\mu$ normalized by $m_\mathrm{PS}$ and the temperature $T$ normalized by $T_c$.

In this paper, we investigate meson masses, a scale setting function and a pseudo-critical temperature by setting the Iwasaki gauge and $N_f=2$ Wilson fermion as lattice actions.
To do this, firstly we measure the meson spectrum and determine a combination of $\beta$ and $\kappa$ that gives a constant mass-ratio between the pseudoscalar and vector mesons. 
Such a combination gives a line  in $\beta$--$\kappa$ plane, so that it is called a line of constant physics. 
Secondly, utilizing the set of lattice parameters on the line of constant physics, we determine a scale setting function.
It gives a relation between the lattice spacings and the lattice bare coupling constant via some reference scale, {\it e.g.} the Sommer scale ($r_0,r_c$)~\cite{Sommer:1993ce, Necco:2001xg} and $t_0$ and $w_0$ scales in the gradient flow method~\cite{Luscher:2009eq, Luscher:2011bx,Luscher:2010iy, Borsanyi:2012zs}.
Here, we take the $w_0$ scale~\cite{Borsanyi:2012zs} as a reference scale.
Furthermore, we estimate a pseudo-critical value of $\beta$  from the chiral susceptibility in the finite-temperature simulations.
Combining the scale setting function with the pseudo-critical temperature on the lattice makes it possible to fix the temperature scale for any value of $\beta$ and $N_\tau$.
Also, it normalizes the axes of the QCD phase diagram to $T/T_c$ and $\mu/m_\mathrm{PS}$.

This paper is organized as follows.
In \S.~\ref{sec:lqc2d}, we explain the lattice setup in our numerical simulations.
In \S.~\ref{sec:obs}, we explain the definition and calculation strategy of observables, which are the meson masses, the $w_0$ scale in the gradient flow method, the Polyakov loop and the chiral condensate.
Section~\ref{sec:T=0} exhibits the results at zero temperature simulation. Firstly we investigate the meson spectrum and then find the line of constant physics. 
Furthermore, we give the scale setting function on the line of constant physics by using the gradient flow method.
In \S.~\ref{sec:finiteT}, we perform the finite temperature simulations and  measure the Polyakov loop and chiral condensate.
From the susceptibility of the chiral condensate, we estimate the pseudo-critical temperature on the lattice.
Section~\ref{sec:summary} is devoted to a summary of this work.

%%%%%%%%%%%%%%%%%%%%%%%%%%%%%%%%%%%%%%%%%%%%%%%%%%%%%%%%%%%%%%%%%%%%%%%%%%%%%%%%%%%%%%%
\section{Lattice action}
\label{sec:lqc2d}

We start with the partition function, which is given by
 $Z=\int {\cal D}U{\cal D}\psi{\cal D}\bar{\psi} e^{-S[U,\psi,\bar{\psi}]}$
 with $U$ the SU(2) link variables,
 $\psi$ ($\bar{\psi}$) the fermions (antifermions) on a lattice,
 and $S$ the Euclidean lattice action given by
 $S[U,\psi,\bar{\psi}]=S_g[U] + S_f[U,\psi,\bar{\psi}]$.
Here $S_g$ and $S_f$ are 
 the actions for the pure gauge and fermion sectors, respectively.

\subsection{Gauge action}
For the SU(2) gauge fields,
 we employ the Iwasaki gauge action
 composed of the plaquette term with $W^{1\times 1}_{\mu\nu}$
 and the rectangular term with $W^{1\times 2}_{\mu\nu}$,  
\beq
S_g = \beta \sum_x \lk
 c_0 \sum^{4}_{\substack{\mu<\nu \\ \mu,\nu=1}} W^{1\times 1}_{\mu\nu}(x) +
 c_1 \sum^{4}_{\substack{\mu\neq\nu \\ \mu,\nu=1}} W^{1\times 2}_{\mu\nu}(x) \rk ,
\eeq
where $\beta=4/g^2$ is the effective coupling constant with $g$ the bare gauge coupling constant,
 and the coefficients $c_0$ and $c_1$ are set to $c_1=-0.331$ and $c_0=1-8c_1$~\cite{Iwasaki:1985we,Iwasaki:2011np}.
Here the plaquette and rectangular Wilson loops are given by
\beq
W^{1\times 1}_{\mu\nu}(x)&=&U_\mu(x)U_\nu(x+\hat{\mu})U_\mu^\dagger(x+\hat{\nu})U_\nu^\dagger(x), \\
W^{1\times 2}_{\mu\nu}(x)&=&U_\mu(x)U_\mu(x+\hat{\mu})U_\nu(x+2\hat{\mu})   
 U_\mu^\dagger(x+\hat{\mu}+\hat{\nu})U_\mu^\dagger(x+\hat{\nu})U_\nu^\dagger(x),
\eeq
 respectively.

\subsection{Fermion action}
For the fermion action, 
 we use the na\"{i}ve two-flavor Wilson fermion action, 
\beq
 S_f = \sum_{f=u,d}\sum_{x,y} \bar{\psi}^f(x) \Delta(x,y) {\psi}^f(y)
\eeq
 with the Dirac operator defined by 
\beq \label{ope}
\Delta(x,y) = \delta_{x,y} - \kappa \sum_{\mu=1}^4
 \ldk \lk 1 - \gamma_\mu \rk U_{x,\mu}\delta_{x+\hat{\mu},y}
 + \lk 1+\gamma_\mu\rk U^\dagger_{y,\mu}\delta_{x-\hat{\mu},y} \rdk ,
 \eeq 
where $\kappa$ is the hopping parameter, $\gamma_\mu$ are the gamma matrices,
 and unless specified explicitly, the lattice spacing $a$ is set to unity, 
 as well as the Wilson parameter $r=1$.
The inverse of (\ref{ope}) corresponds to the fermion propagator,
 while the inverse of $\kappa$ related to the quark mass,
 which reads $1/\kappa=2 m_q^0+8$ with $m_q^0$ the bare quark mass in the tree level on the lattice.
 In the quantum level, by using of the quark mass renormalization constant $Z_m$,
 the hopping parameter is related to the renormalized quark mass as $\kappa=(2Z_m m_q^R+\mbox{const.})^{-1}$,
 where the constant corresponds to an additive renormalization depending on $\beta$.
The value of $\kappa$, where the renormalized quark mass is zero, is called the critical $\kappa$, 
 which is represented by $\kappa_c$.

%%%%%%%%%%%%%%%%%%%%%%%%%%%%%%%%%%%%%%%%%%%%%%%%%%%%%%%%%%%%%%%%%%%%%%%%%%%%%%%%%%%%%%%%
\section{Observables} 
\label{sec:obs}
In order to obtain the expectation values of observables defined by
\beq
 \bra {\cal O} \ket = \frac{1}{Z} \int{\cal D}U {\cal O} \ldk \mbox{det}\Delta(U) \rdk e^{-S[U]},
\eeq
 one needs gauge configurations $U$ beforehand.
To this end,
 we use the hybrid Monte Carlo (HMC) algorithm~\cite{Duane:1987de,Gottlieb:1987mq}
 as a method for generating gauge configurations.
This algorithm combines
 a Metropolis acceptance or rejection step for forming a Markov chain
 with a molecular dynamics evolution for proposing new configurations,
 and is usually employed for dynamical simulations
 including the fermion contributions. 
In the present paper,
 we generate the gauge configurations for $N_f=2$ two-color QCD at vanishing chemical potential.
Using these configurations, 
 we here measure several observables: 
 the meson correlation function, 
 the Yang-Mills action density, 
 the Polyakov loop and the chiral condensate.

\subsection{Meson masses} 
\label{subsec:meson}
For the determination of the line of constant physics,
 we use meson masses
 and consider the pseudoscalar to vector meson mass ratio $m_\mathrm{PS}/m_\mathrm{V}$.
Note that mesons are indistinguishable from baryons in two-color QCD at vacuum
 due to the Pauli-Gurs\"{e}y symmetry~\cite{Pauli:1957,Gursey:1958},
 where baryons are bosonic and the lightest baryon (diquark) becomes a Nambu-Goldstone (pionic) mode.

The meson masses can be extracted from
 the Euclidean time dependence of the meson correlation functions.
The temporal meson correlator projected to zero spatial momentum 
 is defined as the connected two-point correlation function: 
\beq
 \label{corr}
 C(\tau) = \int d^3 x \bra {\cal M}(\tau,\vx) {\cal M}^\dagger(0,\vec{0}) \ket_\mathrm{c} 
\eeq
 with the meson operator given by
\beq
 {\cal M}(x) = \bar{\psi}^{a}_{\alpha}(x) \Gamma_{\alpha\beta} \psi^{a}_{\beta}(x) ,
\eeq
where $\Gamma$ is a product of Dirac matrices
 and roman (greek) letters denote color (spinor) indices.
Here $\Gamma = {\mbox{1}\hspace{-0.25em}\mbox{l}}, \gamma_5, \vec{\gamma}, \gamma_5\gamma_4$
 correspond to the scalar, pseudoscalar (PS), vector (V) and axial vector mesons, respectively.
The full expression for Eq.~(\ref{corr}) reads
\beq
 C(\tau) = \frac{1}{Z} \int {\cal D}U \mbox{det} \ldk \Delta(U) \rdk e^{-S_g[U]}  \int d^3x \mathrm{Tr}_{c,s}
 \lk \Gamma \Delta^{-1}(\tau,\vx, 0,\vec{0}) \Gamma^{\dagger} \Delta^{-1}(0,\vec{0},\tau,\vx) \rk ,
\eeq
 where $\mathrm{Tr}_{c,s}$ stands for the trace in color and spinor space.

The meson correlator includes all energy eigenstates as
$C(\tau) 
 = \sum_n |\bra n | {\cal M} | \Omega \ket|^2 e^{-(E_n -E_\Omega) \tau}$,
 where $|\Omega \ket$ and $E_{\Omega}$ represent the vacuum state and the vacuum energy, respectively.
In the limit $\tau \rightarrow \infty$, only the the smallest energy gap $E_1 - E_\Omega$ remains.
The effective masses for the PS and vector meson are given by the energy gap for each quantum number obtained
by taking  $\Gamma = \gamma_5$ and $\vec{\gamma}$, respectively.

In the numerical lattice simulations, these masses can be read off from the slope of the calculated correlation function. 
Considering the effective mass defined by
\beq \label{effmass}
 m_\mathrm{eff}(\tau) = -\ln \lk C(\tau+1)/C(\tau) \rk ,
\eeq
 and plotting it as a function of $\tau$,
 a plateau appears for sufficiently large $\tau$ to be dominated by the ground state;
 the meson mass is expected to be obtained
 by fitting a function for the mass spectrum to a mass plateau.

When actually computing the meson mass, 
 the meson correlator, which is measured on a finite lattice, has
 the periodic boundary conditions in Euclidean time imposed by using the time period $L_\tau$,
 with the result that the meson propagation becomes symmetric in $\tau$ and $L_\tau-\tau$.
This implies that the meson correlation function cannot be described
 by a single exponential function mentioned above.
Thus, in practice, we fit the meson correlator
 to a sum of two exponential functions, that is, a hyperbolic cosine (cosh) form:
\beq
 C(\tau) = C_0 \mathrm{cosh}\ldk m \lk \tau - L_{\tau}/2 \rk \rdk ,
\eeq
where $C_0$ is the amplitude and $m$ is the mass.
Note here that, on the lattice, this fit equation reads
 $C(\tau/a) = C_0 \mathrm{cosh}\ldk am \lk \tau/a - N_\tau/2 \rk \rdk$
 with the lattice spacing $a$ and the temporal lattice extent $N_\tau=L_\tau/a$.

\subsection{Reference scale for the scale-setting function}
\label{subsec:t2E}
The scale-setting function on the lattice relates the
  lattice spacing $a$
 to the inverse bare gauge coupling $\beta$.
To determine the relationship, we introduce a reference scale since the QCD-like theory has a dynamical scale, $\Lambda_{\mathrm{QCD}}$, which is mainly determined by the gluon dynamics.
We here employ the gradient flow method~\cite{Luscher:2010iy} and utilize the $w_0$ scale proposed in Ref.~\cite{Borsanyi:2012zs}.

The gradient flow evolves the original gauge fields $A_\mu(x)$
 in the direction of an extra dimension $t$ with mass-dimension $-2$,
 which is called flow time, 
 toward local minima (stationary points) of the Yang-Mills action $S_g$. 
The flowed gauge fields $B_\mu(x,t)$ are defined by the following equation~\cite{Luscher:2010iy},
\beq \label{ymgf}
\partial_t B_\mu(x,t) = D_\nu G_{\nu\mu}(x,t) ,
 \quad D_\mu = \partial_\mu + \ldk B_\mu(x,t), \ \cdot \ \rdk,
\eeq
with the initial condition $B_\mu(x,t)|_{t=0}=A_\mu(x)$
 and the field strength tensor on the flowed field given by
\beq
 G_{\mu\nu}(x,t) = \partial_\mu B_\nu(x,t) - \partial_\nu B_\mu(x,t) + \ldk B_\mu(x,t),B_\nu(x,t) \rdk .
\eeq
Here, the gauge fields are smeared over a region of radius $|x|=\sqrt{8t}$.
 The composite operators made of $B_\mu(t)$ are automatically renormalized at positive flow time.
On the lattice, Eq.\ ({\ref{ymgf})} is defined by 
 $\partial_t V_t(x,\mu)=-g^2 [\partial_{x,\mu}S_g(V_t)] V_t(x,t)$ 
 with the initial condition $V_t(x,t)|_{t=0}=U_\mu(x)$~\cite{Luscher:2010iy,Luscher:2009eq},
 where $V_t$ is the flowed link valuable and 
 $\partial_{x,\mu}$ is the SU(2)-valued difference operator.
All quantities calculated from the flowed link valuables depend on the dimensionful flow-time $t$,
 so that one can define a scale at a specific $t$
 where a chosen dimensionless quantity reaches a fixed value.
Such a dimensionless quantity includes, for example,
 the product of the flow time squared and 
 the expectation value of the gauge field energy density (Yang-Mills action density),
 that is, $t^2\bra E(t) \ket$ with $E=G_{\mu\nu}G_{\mu\nu}/4$.
This quantity is finite in the continuum limit~\cite{Luscher:2011bx}, 
 and thus the reference time-scale is independent of the lattice spacing.
For this reason, $t^2\bra E(t) \ket$ is a good candidate for scale determination.

The reference time-scale $t_0$ (sometimes termed the $t_0$-scale) is defined by
\beq
 t^2 \bra E(t) \ket|_{t=t_0} = {\cal T}
\eeq
with ${\cal T}$ the reference value fixed as a fiducial point.
Here, $E$ is a smeared local operator, so that the use of the $t_0$-scale leads to small statistical uncertainties.
Empirically, with increasing $t$,
 the energy density $t^2\bra E(t) \ket$ rises rapidly from $0$
 and then becomes a linear function~\cite{Luscher:2010iy,Borsanyi:2012zs}.
${\cal T}$ is a constant and its value should be chosen to be a certain value in this linear regime,
 and also be taken so as to reduce discretization errors and to suppress finite-volume effects.
For smaller ${\cal T}$ the discretization errors become larger,
 while for larger ${\cal T}$ statistical errors grow because of a large value of $t$ together with numerical costs.
In the case of ${\cal T}=0.3$ for the SU(3) gauge theory,
 the $t_0$-scale shows a comparable scale with the Sommer reference scale~\cite{Sommer:1993ce,Luscher:2010iy} 
 (see~\cite{Sommer:2014mea} for ${\mathcal T}=2/3$). 
For the SU(2) gauge theory, on the other hand,
 it has been recently reported that
 the $t_0$-scale with ${\cal T}=0.1$ is comparable with the Necco-Sommer scale~\cite{Necco:2001xg,Hirakida:2018uoy}
 (see also~\cite{Berg:2016wfw,Giudice:2017dor} for different values of ${\cal T}$). 

We here consider an alternative reference scale, termed as the $w_0$-scale~\cite{Borsanyi:2012zs}.
$w_0$ is defined by
\beq
 t \frac{d}{dt} t^2 \bra E(t) \ket|_{t=w_0^2} = {\cal W}.
\eeq
Although the $t_0$-scale is defined by $t^2\langle E \rangle$, which incorporates information about the link variables from all scale larger than $\mathcal{O} (1/\sqrt{t})$,
the $w_0$-scale depends on the scales only around $\sim 1/\sqrt{t}$.
Thus, the $t_0$-scale suffers from the larger discretization errors coming from the data in a small flow-time regime ($t \lesssim a^2$).
We mainly take a reference value ${\cal W}=0.3$, and also choose ${\cal W}=0.5$ to obtain the scale-setting function for the wider $\beta$ regime.

\subsection{Polyakov loop and chiral condensate} 
\label{subsec:Ploop}
By analyzing an appropriate order parameter for the phase transition in two-color QCD,
 one can find the critical temperature $T_c$.
This critical value is determined by looking for a peak in the susceptibility of the order parameter.

The Polyakov loop is known
 as an approximate order parameter of the confinement-deconfinement phase transition;
 this acts as an exact order parameter for spontaneously center symmetry breaking
 in pure Yang-Mills theory (or a system where the quark mass is infinitely large).
For SU(2) gauge theory, the Polyakov loop is defined by 
\beq
 L = \frac{1}{N_s^3}\sum_{\vx}\frac{1}{2}\mathrm{tr}\prod_{\tau=1}^{N_\tau} U_4(\vx,\tau) \label{eq:def-Ploop}
\eeq
 with the SU(2) link valuables $U_\mu$ and the spatial lattice size $N_s$.
The ensemble average of the expectation value of the Polyakov loop $L$
 is related to the free energy of a single probe quark (antiquark): 
 $\langle L \rangle \sim e^{-F_q/T}$.
This implies that $\langle L \rangle =0$ in the confined phase due to $F_q \rightarrow \infty$,
 while in the deconfined phase $\langle L \rangle \neq 0$ with a finite value of $F_q$. 
The Polyakov loop susceptibility is defined by
\beq
 \chi_L = {N_s^3} \ldk \bra L^2 \ket - \bra L \ket^2 \rdk . \label{eq:def-chiP}
\eeq

Although the confinement-deconfinement transition associated with the center symmetry is of second order in the SU(2) gauge theory, 
 once we introduce dynamical fermions to the system 
 the transition alters to the crossover 
due to the explicitly broken center symmetry.
As for the chiral phase transition,  on the other hand,  
one can define the phase transition in the chiral limit, 
for chiral symmetry even though the system includes the dynamical fermions.
In this work, we utilize the Wilson fermion on the lattice, and then both the center and chiral symmetries are explicitly broken.
Therefore, we expect that there is no phase transition, but a crossover occurs.
We measure both order parameters, namely, the Polyakov loop and chiral condensate, and estimate the pseudo-critical temperature, $T_c$, of the chiral phase transition.

Next, we consider the chiral condensate with Wilson fermions.
Because of the Wilson term,  the chiral symmetry is explicitly broken, and an appropriate subtraction is needed to correctly obtain the continuum limit. 
The subtracted chiral condensate 
 is given by the following expectation value~\cite{Bochicchio:1985xa,Itoh:1986gy,Aoki:1997fm, Giusti:1998wy,Umeda:2012nn,Hayakawa:2013maa,Iritani:2015ara}:
\beq \label{subchi}
\bra \bar{\psi}^f \psi^f \ket_\mathrm{sub}
 = 8 \kappa^2 m_\mathrm{PCAC} 
 \sum_x \bra {\cal M}_\mathrm{PS}(\tau,\vx) {\cal M}_\mathrm{PS}^\dagger(0,\vec{0}) \ket \label{eq:def-chiral}
\eeq
with the partially conserved axial current (PCAC) mass $m_\mathrm{PCAC}$ defined by
\beq \label{eq:pcac}
m_\mathrm{PCAC} 
 = \frac{\sum_{\vx}\partial_4 \bra {\cal M}_{\mathrm{A}_4}(\tau,\vx) {\cal M}_{\mathrm{PS}}^\dagger(0,\vec{0})\ket}
   {2\sum_{\vx} \bra {\cal M}_{\mathrm{PS}}(\tau,\vx){\cal M}_{\mathrm{PS}}^\dagger(0,\vec{0}) \ket},
\eeq
via the Ward-Takahashi identity.
Here, ${\cal M}_\mathrm{PS}$ and ${\cal M}_{\mathrm{A}_\mu}$ denote
 the pseudoscalar and axial vector currents
 defined by ${\cal M}_\mathrm{PS}=\bar{\psi}^f \gamma_5 \psi^f$
 and ${\cal M}_{\mathrm{A}_\mu}=\bar{\psi}^f \gamma_5\gamma_\mu \psi^f$, respectively, and $m_\mathrm{PCAC}$ is the value at zero temperature.
We do not care about the flavor-mixing condensate, and thus only consider flavor-diagonal one.
Also, the label of flavor, $f$, is fixed.
The corresponding susceptibility is given by 
\beq
\chi_\mathrm{chi} 
 = \bra {\cal O}_\mathrm{chi}^2 \ket - \bra {\cal O}_\mathrm{chi} \ket^2  \label{eq:def-chi}
\eeq
with
\beq
{\cal O}_\mathrm{chi}
 = 8 \kappa^2 m_\mathrm{PCAC} \sum_x {\cal M}_\mathrm{PS}(\tau,\vx) {\cal M}_\mathrm{PS}^\dagger(0,\vec{0}) . 
\eeq
We call this here the ``chiral susceptibility,''
 which has mass dimension $6$.
While this definition is different from that for the conventional chiral susceptibility $\chi$
 with mass dimension $2$ (e.g., see~\cite{Fodor:2009ax}),
 it is related via the dimensionless value: $\chi/T^2 = N_s^3 N_\tau^3 \chi_\mathrm{chi}$.

%%%%%%%%%%%%%%
%%    results
%%%%%%%%%%%%%%
\section{Results at zero temperature}
\label{sec:T=0}
We here perform zero temperature simulations
 on the lattices with 
 $N_s^3 \times N_\tau = 16^3 \times 32$
 using the Iwasaki gauge action and the $N_f=2$ Wilson fermion action.
 We determine the line of constant physics and the scale together with the mass renormalization. 
Firstly, we have generated the gauge configurations for various combinations of ($\beta,\kappa$) at $T=0$ and $\mu=0$ by using the HMC method.
For each parameter set of ($\beta, \kappa$), we sampled one configuration every $20-100$ trajectories
 and eventually produced about $30-150$ configurations,
 where we discarded  some configurations corresponding to $200-250$ trajectories
 at the beginning of the Monte Carlo trajectories due to thermalization.

\subsection{Line of constant physics} 
To determine the line of constant physics, we take a reference value, $m_{\mathrm{PS}}/m_\mathrm{V}=0.8232$,
which corresponds to the result at $\beta=0.80, \kappa=0.1590$.
The meson masses are extracted from the meson correlation functions.
We calculate the meson correlation function for various values of $\kappa$ for each $\beta$,
 using the Coulomb gauge-fixing and the wall sources.

%%%%%%%%%%%%%%%%
\begin{figure*}[h]
\centerline{\includegraphics[scale=0.7]{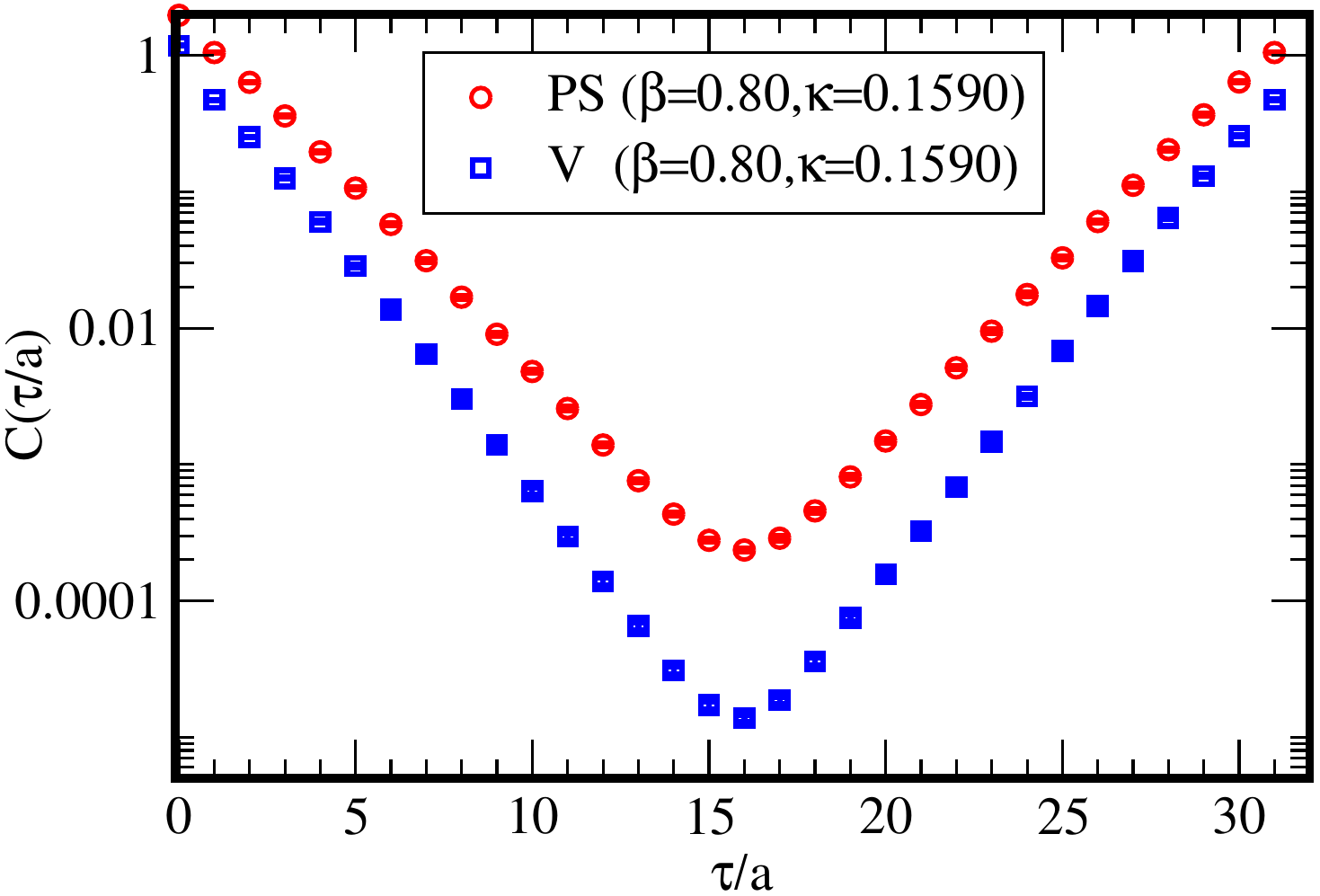}}
\caption{Correlation functions for the pseudoscalar (PS) and vector (V) mesons at ($\beta,\kappa$) $=$ ($0.80,0.1590$).
}\label{fig:mcorr}
\end{figure*}
%%%%%%%%%%%%%%%%
Figure \ref{fig:mcorr} shows the correlation functions in the PS and vector meson channels  at ($\beta,\kappa$) $=$ ($0.80,0.1590$).
When extracting effective masses,
 instead of a simple ratio $C(\tau/a+1)/C(\tau/a)$ in Eq.~(\ref{effmass}),
 we utilized an alternative one that has
 not only the forward slope but also the backward slope of the correlator 
taken into account
 to achieve better accuracy:
 $[C(\tau/a+1)+C(\tilde{\tau}/a-1) ]/[ C(\tau/a)+C(\tilde{\tau}/a)]$,
 where the forward imaginary time in lattice units $\tau/a$ and the backward one $\tilde{\tau}/a$
 are symmetric about $N_\tau /2$.
The fitting is performed by using only larger time slices
 to omit the contributions of excited modes.

%%% Fig.2 %%%%%%%%%%%%%%%%%%%%%%%
\begin{figure*}[h]
\centerline{\includegraphics[scale=0.7]{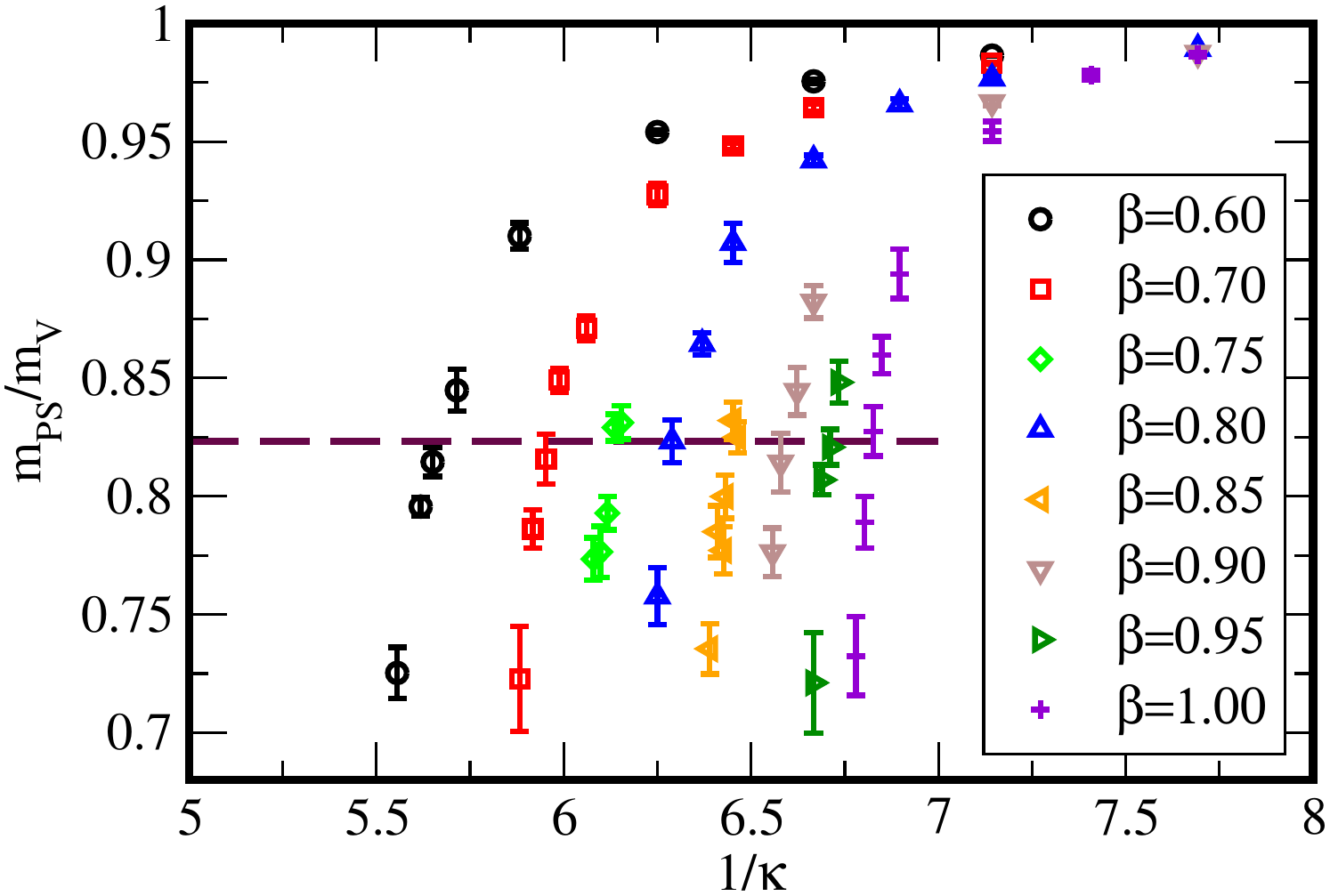}}
\caption{$m_\mathrm{PS}/m_\mathrm{V}$ for various combinations of the hopping parameter $\kappa$ and $\beta$.}\label{fig:mps-mv}
\end{figure*}
%%%%%%%%%%%%%%%%%%%%%%%%%%%%
%%% Tab.1 %%%
\begin{table}[tbp]
\centering
\begin{tabular}{ccc}
  \hline \hline 
  $\beta$ & $\kappa (m_\mathrm{PS}/m_\mathrm{V}=0.8232)$ & $\kappa_c$ \\
  \hline
   0.60 & 0.1764 & 0.1839(1) \\ 
   0.70 & 0.1678 & 0.1728(1) \\ 
   0.75 & 0.1629 & 0.1673(1) \\ 
   0.80 & 0.1590 & 0.1627(1) \\ 
   0.85 & 0.1549 & 0.1584(1) \\    
   0.90 & 0.1515 & 0.1546(0) \\ 
   0.95 & 0.1490 & 0.1515(1) \\ 
   1.00 & 0.1465 & 0.1489(1) \\ 
 %  1.05 & 0.1450 &  \\ 
  % 1.10 & 0.1437 & 0.1453(1) \\ 
  \hline \hline
\end{tabular}
\caption{\label{tab:lat-param}
 Summary of the values of $\kappa$ consistent with $m_\mathrm{PS}/m_\mathrm{V}=0.8232$
 and $\kappa_c$ for each $\beta$.}
\end{table}
%%%%%%%%%%%%%%%%%%%%%%

For details of the resulting masses, we refer the reader to tables \ref{app:masses1} and \ref{app:masses2} in the Appendix.
In figure~\ref{fig:mps-mv}, we show the $m_\mathrm{PS}$-to-$m_\mathrm{V}$ ratio
 as a function of the inverse hopping parameter $1/\kappa$ for each $\beta$.
 We determine the value of $\kappa$  consistent with $m_\mathrm{PS}/m_\mathrm{V}=0.8232$ for each $\beta$     
 by linearly interpolating $3$ data near $m_\mathrm{PS}/m_\mathrm{V}=0.8232$.
The results for ($\beta,\kappa$)  are shown in table \ref{tab:lat-param}.
Here, the statical errors of the $\kappa$ values
on the line of $m_\mathrm{PS}/m_\mathrm{V}=0.8232$ are to be less than $2\%$ accuracy.

\subsection{Mass renormalization} 
\begin{figure*}[h]
\centerline{\includegraphics[scale=0.5]{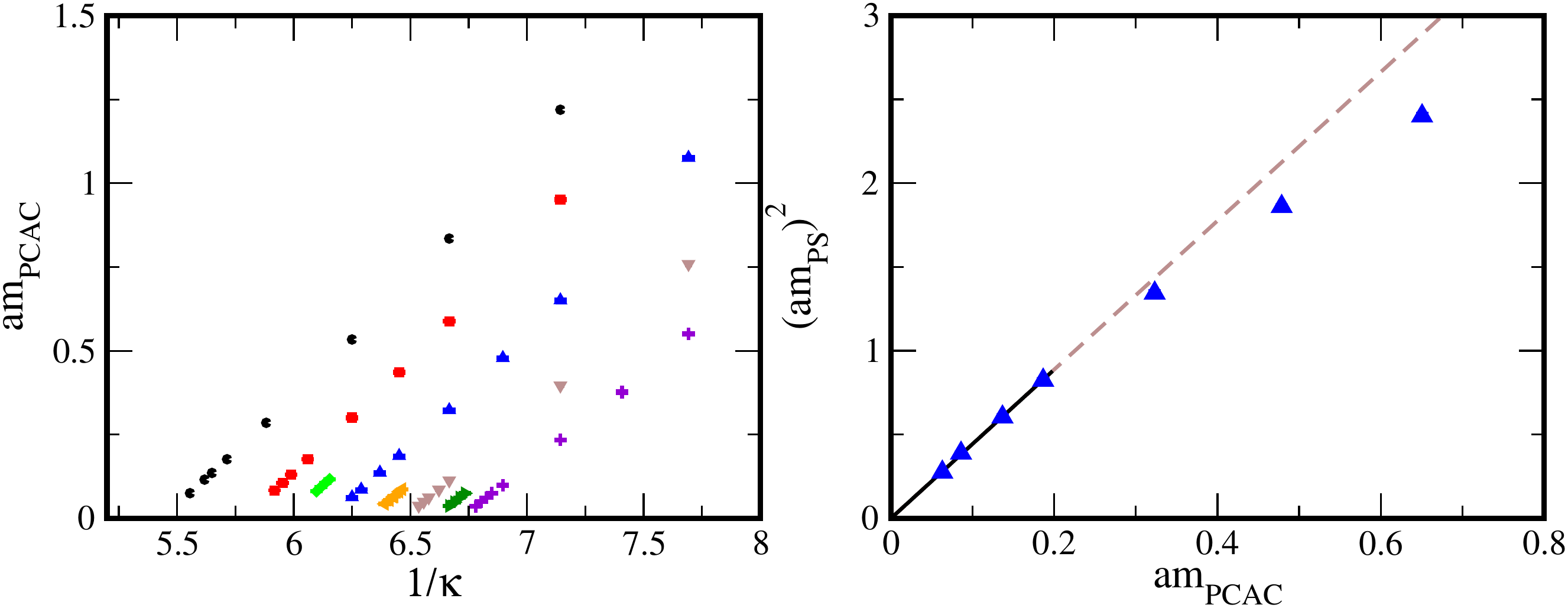}}
\caption{(Left) The hopping parameter dependence of $m_\mathrm{PCAC}$.
Each color denotes the same value of $\beta$ as in Fig.~\ref{fig:mps-mv}.
%Each line denotes the fit result for smaller values of $m_\mathrm{PCAC}$
(Right) 
The PCAC mass dependence of the PS meson mass squared at $\beta=0.80$. The dashed line denotes the linear fitting-function by which we fit four data points in the small $am_\mathrm{PCAC}$ regime.}\label{fig:mpcac}
\end{figure*}

The PCAC mass can be read off from Eq.~(\ref{eq:pcac})
 using the calculated meson correlation functions.
Figure \ref{fig:mpcac} shows the PCAC mass as a function of $1/\kappa$ at fixed $\beta$,
 which gives the relation between the renormalized mass and bare mass of the quark.
Here, for the difference operator in Eq.~(\ref{eq:pcac}),
 we have considered not only the forward difference $\partial_4$ but also the backward difference $\tilde{\partial}_4$
 (i.e., $\partial_4 \rightarrow \frac{1}{2}(\partial_4+\tilde{\partial}_4)$)
 to achieve better accuracy.
For smaller values of $m_\mathrm{PCAC}$,
 we can fit them with the function of  $(1/\kappa -1/\kappa_c)$.
 We find the extrapolated value for $1/\kappa$ at $m_\mathrm{PCAC}=0$,
 which leads to $\kappa_c$.
The values of $\kappa_c$ for each $\beta$ are summarized in table \ref{tab:lat-param}.

%%%%%%%%%%%%%%%%%
%%%%%%%%%%%%%%%%%
%%%%%%%%%%%%%%%%%
%%%%%%%%%%%%%%%%%

We also present the PS meson mass squared as a function of the PCAC mass as exemplary data ($\beta=0.80$) in lattice units in the right panel of Fig.~\ref{fig:mpcac}.
 We can see that $m^2_\mathrm{PS}$ linearly depends on $m_\mathrm{PCAC}$
 in the regime of smaller masses:
\beq
 a^2 m_\mathrm{PS}^2 \propto a m_\mathrm{PCAC} ,
\eeq
 which implies that the PCAC relation (or the Gell-Mann--Oakes--Renner relation for the PS meson mode) holds.
Thus, $m_\mathrm{PS}\rightarrow 0$ in the limit of $m_\mathrm{PCAC} \rightarrow 0$.

Note that in two-color QCD it is known that the chiral symmetry can be restored in the chiral limit even at zero chemical potential.
Instead of the spontaneous breaking of the chiral symmetry, the spontaneous breaking of the baryon charge symmetry occurs because of meson-baryon symmetry in two-color QCD~\cite{Kogut:1999iv, Furusawa:2020qdz}.

\subsection{Scale-setting function} 
Now we determine the relative scale of $a$ for several $\beta$ using the reference scale $w_0$ in the gradient flow method.
First of all, we have generated the gauge configurations on $16^3 \times 32$ on the line of constant physics.
The resulting meson spectrum from the generated configurations is summarized in Table \ref{tab:mass-param}.
%%% Tab.2 %%%
\begin{table}[h]
\centering
\begin{tabular}{cccc}
  \hline \hline 
  $\beta (\kappa)$ & $am_\mathrm{PS}$ & $ m_\mathrm{PS}/m_\mathrm{V}$ & $a m_\mathrm{PCAC}$ \\
  \hline
   0.60 (0.1764) & 0.9075(11) & 0.8336(50) & 0.1482(3) \\ 
   0.70 (0.1678) & 0.7655(23) & 0.8217(88) & 0.1107(7) \\ 
   0.75 (0.1629) & 0.7216(29) & 0.8260(49) & 0.1073(5) \\ 
   0.80 (0.1590)  & 0.6229(34) & 0.8233(90) & 0.08593(77) \\ 
   0.85  (0.1549) & 0.5748(29) & 0.8179(80) & 0.08240(62) \\    
   0.90 (0.1515) & 0.5092(38) & 0.8118(50) & 0.07299(66) \\ 
   0.95 (0.1490) & 0.4349(42) & 0.8207(75) & 0.06206(60) \\ 
   1.00 (0.1465) & 0.4108(58) & 0.8274(105) & 0.06159(94) \\ 
  \hline \hline
\end{tabular}
\caption{\label{tab:mass-param}
 Summary of the values of $m_\mathrm{PS}$, $m_\mathrm{PS}/m_\mathrm{V}$, and $m_\mathrm{PCAC}$ 
 for each $\beta$ on the line of constant physics.}
\end{table}
%%%%%%%%%%%%%%%%%%%%%%
Next, we have numerically solved the gradient flow by starting with the generated configurations and utilizing the
Wilson gauge action for the gradient flow process.
Finally, we have measured the expectation value of $t^2 \bra E(t) \ket$  in the range of $\beta=0.60-1.00$ using the clover-leaf definition for $G_{\mu \nu}$.
Here, we have estimated the statistical errors using the Jackknife method.
%%%%%%%%%%%%%%%%%%%%%%
%%%%%%%%%%%%%%%%%%%%%%
\begin{figure*}[h]
\centerline{\includegraphics[scale=0.5]{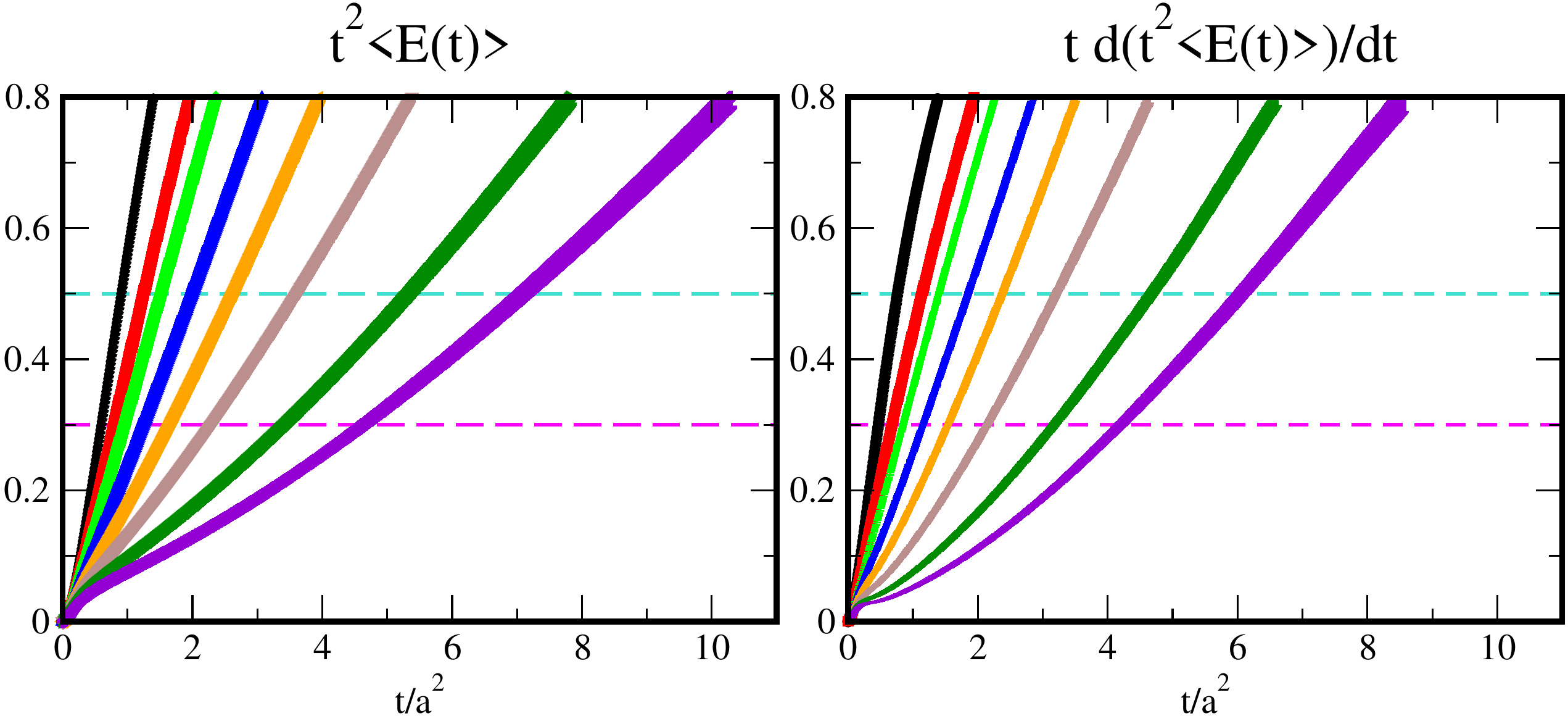}}
\caption{\label{fig:t0}
Results for the dimensionless combination $t^2\langle E(t) \rangle$ and $t d(t^2 \langle E(t) \rangle)/ dt$
 as a function of the flow time in lattice units $t/a^2$.
We plot the data for $0.60 \le \beta \le 1.00$ and each color denotes the same value of $\beta$ as in Fig.~\ref{fig:mps-mv}.
Two dashed lines, magenta and cyan, denote ${\cal T}$ or ${\cal W}=0.30$ and $0.50$, respetively.}
\end{figure*}
%%%%%%%%%%%%%%%%%%%%%%
%%%%%%%%%%%%%%%%%%%%%%

Figure \ref{fig:t0} shows the results for $t^2\bra E(t) \ket$ and $t d t^2 {\bra E(t)}/dt$ at each $\beta$.
Here, ${\cal T}=0.30, 0.50$ and ${\cal W}=0.30, 0.50$ are also plotted as guiding lines of the reference scales in the left and right panels, respectively.
We see that all reference scales for ${\cal T}$ and ${\cal W}$ are located in the approximately linear-scaling region of the observables.

%%%  %%%
\begin{table}[h]
  \centering
  \begin{tabular}{c|c|c|}
% \hline   
 &  ${\cal W}=0.30$ & ${\cal W}=0.50$  \\
  \hline
$\beta (\kappa) $  & $w_0/a$   & $w_0/a$  \\  
0.60 (0.1764) &   0.6830(4)    &  0.8733(12)    \\
0.70 (0.1678) &   0.8219(15)   &  1.067(2)      \\
0.75 (0.1629) &   0.9205(21)   &  1.182(4)      \\
0.80 (0.1590) &    1.074(4)    &  1.360(6)      \\
0.85  (0.1549) &    1.240(7)    &  1.556(11)    \\
0.90 (0.1515) &    1.463(16)   &  1.791(20)    \\
0.95 (0.1490) &    1.781(29)   &  2.162(40)    \\
1.00 (0.1465) &    2.048(40)   &  2.462(59)   \\
%  \hline
  \end{tabular}
  \caption{
The relation between $\beta$ and $w_0/a$ at ${\cal W}=0.30, 0.50$.}
\label{tab:t0}
\end{table}
%%%%%%%%%%%%%%%%%%%

Table~\ref{tab:t0} gives the value of $w_0/a$ for each $\beta$ at ${\cal W}=0.30$ and ${\cal W}=0.50$.
To obtain the scale setting function around $T_c$, we utilize the smaller reference scale ${\cal W}=0.30$, which gives less statistical uncertainty, and find 
\beq
\log(w_0/a) = -0.6469 -1.045\beta + 2.428 \beta^2.
\eeq
Here, we utilize $0.80 \le \beta \le 1.00$ for the interpolation.
To see the wider region of $\beta$, taking the lager reference scale ${\cal W}=0.50$ is better to avoid  strong discretization errors that are inevitable for $w_0 \ll a$.
The  best-fit function is expressed by
\beq
\log(w_0/a) = -0.3822  -0.9212 \beta +2.221\beta^2
\eeq
for $0.60 \le \beta \le 1.00$.

\section{Results at finite temperature} \label{sec:finiteT}

We perform finite temperature simulations
 on the lattice with the size $N_s=32$ and $N_\tau= 10$
 using the same lattice action as the $T=0$ calculations. 
To determine the critical temperature on the line of constant physics, we set the parameter set of ($\beta, \kappa$) as shown in Table~\ref{tab:lat-param}. 
 We have sampled one configuration every $10$ trajectories
 and eventually produced $600$ configurations for each $\beta$. 
We observe the Polyakov loop and chiral condensate as well as the corresponding susceptibilities.
Finally, we will estimate the pseudo-critical temperature from a peak position of the chiral susceptibility.

\begin{figure*}[h]
\centerline{\includegraphics[scale=0.5]{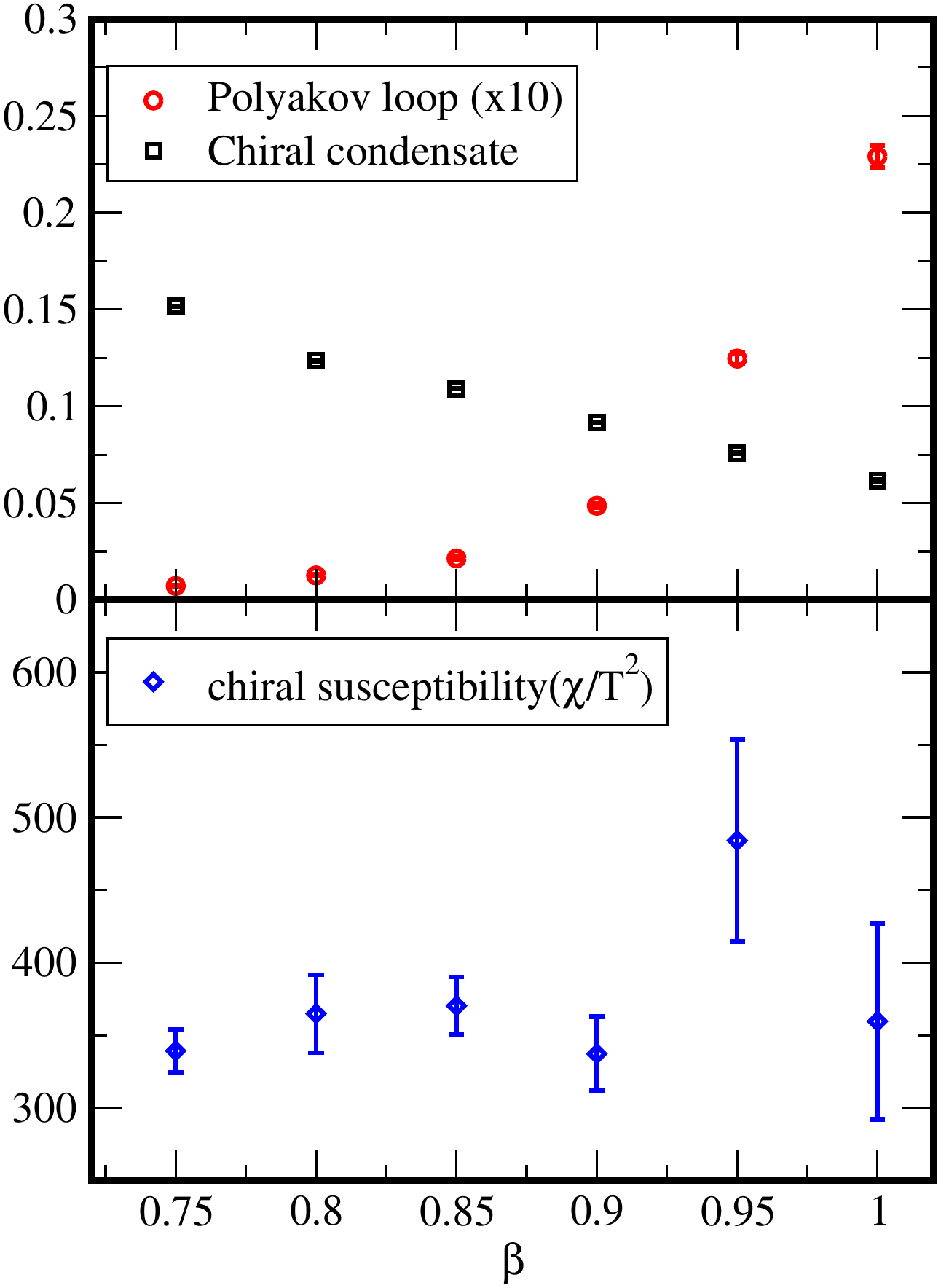}}
\caption{
(Top) The Polyakov loop and the subtracted chiral condensate as a function of $\beta$. The lattice size is $32^3 \times 10$.
For easy seeing, we display the value of the Polyakov loop multiplied by a factor $10$.
(Bottom) The chiral susceptibility as a function of $\beta$. }\label{fig:Ploop-chiral}
\end{figure*}

Following Eqs.~(\ref{eq:def-Ploop}), (\ref{eq:def-chiP}), (\ref{eq:def-chiral}) and (\ref{eq:def-chi}), we calculate the Polyakov loop, the susceptibility of the Polyakov loop, the subtracted chiral condensate and the chiral susceptibility, respectively.
Note here that we use $a m_\mathrm{PCAC}$ for each $\beta$ in Table~\ref{tab:mass-param}
 in calculating the subtracted chiral condensate (\ref{subchi}).
The top panel of Fig.~\ref{fig:Ploop-chiral} depicts the Polyakov loop (red-circle) and the subtracted chiral condensate (black-square), while the bottom panel shows the chiral susceptibility on $32^3 \times 10$ lattice.
By scanning a range of $\beta=0.75-1.00$, we find that the Polyakov loop increases while the chiral condensate decreases in this regime.
Although the susceptibility of the Polyakov loop is very noisy and we cannot find its peak, we can see a peak of the chiral susceptibility.

%%%%%%%%%%%%%%%%%
%%%%%%%%%%%%%%%%%
%%%%%%%%%%%%%%%%%

A peak at $\beta=0.95$ is shown in the bottom panel of Fig.~\ref{fig:Ploop-chiral}.
Here, we identify the value as a pseudo-critical $\beta$, i.e., $\beta_c$.
It is very hard to determine the precise value of $\beta_c$ since we have to fix a value of $\kappa$ for each $\beta$ in such a way as to reproduce the reference value of $m_\mathrm{PS}/m_\mathrm{V}$.

We have also calculated the same quantities on $32^3 \times 6$ and $32^3 \times 8$.
In these cases, however, we cannot find a clear peak of the susceptibility.
The difficulty of the determination of $\beta_c$ might come from the weakness of the phase transition in two-color QCD.
As for the order of phase transition in the SU(3) gauge theory, namely, quenched QCD, the chiral phase transition for probe quarks and color confinement, which are known to simultaneously occur, are of first-order.
In $N_f=2$ massless three-color QCD, on the other hand,
there remains a controversy about whether the chiral phase transition is described by the first- or second-order phase transition~\cite{Pisarski:1983ms, Aoki:2012yj}.
The SU(2) gauge theory or quenched two-color QCD is governed by the second-order phase transition~\cite{Hirakida:2018uoy, Berg:2016wfw, Giudice:2017dor}, since the smaller number of degrees of freedom weakens the phase transition.
We expect that the $N_f=2$ two-color QCD undergoes a further weak phase transition or crossover even in the massless limit and that the  peak of susceptibility must be even broader because of finite mass effects.

Finally, we translate the lattice parameters to the temperature normalized by the pseudo-critical temperature.
The temperature on the lattice is defined by the inverse of the lattice extent in the Euclidean time direction,
\beq
 T = \frac{1}{N_\tau a} .
\eeq
We identify the pseudo-critical value as $\beta_c=0.95$ and $N_\tau=10$ by looking for the peak in the chiral susceptibility.
Thus, we define the pseudo-critical temperature  as 
$ T_c = \frac{1}{N_{\tau} a(\beta_c)}$.
This definition of the pseudo-critical temperature includes the finite-volume effects, so that ideally we must study the thermodynamic limit $N_s \rightarrow \infty$ and continuum limit $N_\tau \rightarrow \infty$.
In the present work, we take the same order of $N_\tau$ as $N_s$ and take this $T_c$ determined by the simulation in $N_\tau=10$ as the pseudo-critical temperature.

Combining with the scale setting function or Table~\ref{tab:t0}, which gives the relation between $a$ and $\beta$, we can calculate $T/T_c$ for any values of $\beta$ and $N_\tau$.
For instance, $\beta=0.80, N_\tau=8$ and $\beta=0.80, N_\tau=16$, where we have investigated the phase structure in the previous work~\cite{Iida:2019rah}, correspond to $T= 0.79(2)(3)T_c $ and $T=0.39(1)(2)T_c$, respectively.
Here, the first bracket denotes the statistical uncertainty coming from the statistical error of the $w_0/a$, while the second one shows the systematic uncertainty coming from the choice of the reference scale ${\mathcal W}=0.30$ or $0.50$.
As we explained, there is unknown uncertainty from the determination of  $\beta_c$, which will be addressed in future works.

%%%%%%%%%%%%%%%
%    Summary
%%%%%%%%%%%%%%%
\section{Summary}\label{sec:summary}
In this paper, we have calculated the meson spectrum in two-color QCD using the Iwasaki gauge and $N_f=2$ Wilson fermion actions on $16^3 \times 32$ lattice.
We have reported a set of lattice parameters ($\kappa, \beta$) on the line of constant physics that satisfies $m_\mathrm{PS}/m_\mathrm{V}=0.8232$.
Furthermore, we have obtained the scale setting function, which gives a relation between the lattice bare coupling constant $\beta$ and the lattice spacing, on the line of constant physics.
To find the scale setting function, we have utilized the $w_0$-scale in the gradient flow method.
It has given a precise scale setting function.
We have also found the pseudo-critical temperature from the chiral susceptibility on $32^3 \times 10$.

Using the results for the scale setting function and the pseudo-critical temperature, which is given by $\beta=0.95, N_\tau=10$, we can finally calculate the temperature scale for any $\beta$ and $N_\tau$ in $0.60 \le \beta \le 1.00$.
The temperature normalized by $T_c$ and the chemical potential normalized by $m_\mathrm{PS}$ would be helpful for a qualitative comparison among several proposed two-color QCD phase diagrams using different lattice actions~\cite{Hands:2006ve, Hands:2010gd, Cotter:2012mb, Makiyama:2015uwa, Braguta:2016cpw, Holicki:2017psk, Boz:2019enj, Iida:2019rah}.
Furthermore, it might be useful to see a comparison between the two-color and three-color QCD phase diagrams.

\appendix
\section{Summary of masses}

\begin{table}[h]
\centering
\begin{tabular}{ccccclc}
  \hline \hline 
  $\beta$ & $\kappa$ & $a m_\mathrm{PS}$ & $m_\mathrm{PS}/m_{\mathrm{V}}$ & $a m_{\mathrm{PCAC}}$ & $\#$ of conf. & $\#$ of trj.  \\
  \hline
  0.6000 & 0.1400 &   2.015(1) &  0.9864(2)   &  1.219(4)  & 30 & 3000 \\
  0.6000 & 0.1500 &   1.767(1) &  0.9755(7)   &  0.8348(20)  & 30 & 3000 \\
  0.6000 & 0.1600 &   1.499(2) &  0.9540(9)   &  0.5336(26)  & 30 & 3000 \\
  0.6000 & 0.1700 &   1.184(2) &  0.9101(56)  &  0.2850(13) & 30 & 3000 \\
  0.6000 & 0.1750 &   0.9751(35) &  0.8448(88)  &  0.1762(17) & 30 & 3000 \\
  0.6000 & 0.1770 &   0.8755(17) &  0.8145(61)  &  0.1354(3)  & 150 & 3000  \\ 
  0.6000 & 0.1780 &   0.8204(15) &  0.7956(40)   & 0.1155(5)  & 300  & 6000  \\ 
  0.6000 & 0.1800 &   0.6805(18) &  0.7252(108) &  0.07519(67) & 300  & 6000  \\ 
  \hline                                               
  0.7000 & 0.1400  &  1.832(1) &  0.9832(4)    &  0.9513(24)  & 30 & 3000 \\ 
  0.7000 & 0.1500  &  1.533(1) &  0.9643(16)    &  0.5875(15)  & 30 & 3000  \\ 
  0.7000 & 0.1550 &  1.366(3) &  0.9481(13)   &  0.4360(16)  & 30 & 3000 \\ 
  0.7000 & 0.1600  &  1.173(3) &  0.9277(47)   &  0.3005(23) & 30 & 3000  \\ 
  0.7000 & 0.1650 &  0.9386(21) &  0.8711(52)   & 0.1763(9)   & 120  & 6000 \\
  0.7000 & 0.1670 &  0.8236(30) &  0.8491(50)   &  0.1301(7)  & 100   & 5000 \\ 
  0.7000 & 0.1680 & 0.7490(16) &  0.8157(106)  &  0.1056(8)  & 90   & 4500 \\ 
  0.7000 & 0.1690 & 0.6734(24) &  0.7861(81)  & 0.08360(53)  & 120 & 6000 \\ 
    0.7000 & 0.1700 &  0.5905(30) &  0.7227(222)  &  0.06222(91)  & 120   & 6000 \\
  \hline                                               
  0.7500 & 0.1625  &  0.7527(15)& 0.8311(79)   & 0.1165(7)  & 170 & 8500  \\
  0.7500 & 0.1630  & 0.7165(14) &  0.8291(56)  &  0.1053(7)  & 120  & 6000 \\
  0.7500 & 0.1635  & 0.6761(12) &  0.7928(70)  &  0.09323(77)  & 180 & 9000 \\
  0.7500 & 0.1640  & 0.6345(28) &  0.7764(108)  &  0.08071(78)  & 120  & 6000 \\
  0.7500 & 0.1645  & 0.5833(20) &  0.7734(89)  &  0.06762(139)  & 120  & 6000 \\
  \hline                                               
  0.8000 & 0.1300 &  1.887(2) &  0.9893(3)    &  1.076(3)  & 30 & 3000 \\
  0.8000 & 0.1400 &  1.550(3) &  0.9768(8)    &  0.6506(30)  & 30 & 3000 \\
  0.8000 & 0.1450 &  1.364(2) &  0.9658(23)   &  0.4785(18)  & 30 & 3000 \\
  0.8000 & 0.1500 &  1.159(3) &  0.9422(20)   &  0.3231(23)  & 30 & 3000 \\
  0.8000 & 0.1550 &  0.9076(32) &  0.9072(83)   &  0.1866(16)  & 30 & 3000 \\
  0.8000 & 0.1570 &  0.7773(48) &  0.8644(48)   &  0.1367(17)  & 60  & 3000 \\
  0.8000 & 0.1590 &  0.6229(34) &  0.8233(90)  &  0.08593(77)  & 120  & 6000 \\
  0.8000 & 0.1600 &  0.5240(32) & 0.7577(120)  &  0.06302(72) & 130 & 6500 \\ 
  \hline \hline
\end{tabular}
\caption{Summary of pseudoscalar (PS), vector (V) and partially conserved axial current (PCAC) masses
 for $\beta = 0.60, 0.70, 0.75, 0.80$, using $16^3 \times 32$ lattice.
The total numbers of configurations and of trajectories are also shown.}
 \label{app:masses1}
\end{table}

\begin{table}[h]
\centering
\begin{tabular}{ccccclc}
  \hline \hline 
 $\beta$ & $\kappa$ & $a m_\mathrm{PS}$ & $m_\mathrm{PS}/m_{\mathrm{V}}$ & $a m_{\mathrm{PCAC}}$ & $\#$ of conf. & $\#$ of trj.  \\
  \hline     
  0.8500 & 0.1547  & 0.5900(24)& 0.8248(66)   & 0.08688(45)  & 140 & 7000  \\                                          
  0.8500 & 0.1550  & 0.5648(33) &  0.8320(79)  &  0.07935(71)  & 120 & 6000 \\    
  0.8500 & 0.1555  & 0.5095(32) &  0.7998(92)  &  0.06585(97) & 120 & 6000 \\ 
  0.8500 & 0.1556  & 0.5062(36) &  0.7770(100)  &  0.06431(56)  & 120  & 6000 \\ 
  0.8500 & 0.1560  & 0.4640(18) &  0.7850(109)  &  0.05459(62)  & 200  & 10000 \\ 
  0.8500 & 0.1565  & 0.4158(66) &  0.7354(106)  &  0.04306(115) & 120 & 6000 \\ 
  \hline                                               
  0.9000 & 0.1300  &  1.591(3) &  0.9874(4)    &  0.7585(32)  & 30  & 3000 \\ 
  0.9000 & 0.1400  &  1.191(3) &  0.9666(16)    &  0.3959(19) & 30 & 3000  \\ 
  0.9000 & 0.1500  &  0.6313(44) &  0.8822(69)   &  0.1128(17) & 30 & 3000 \\  
  0.9000 & 0.1510  &  0.5553(35) &  0.8444(101)   &  0.08537(116) & 60 & 3000 \\ 
  0.9000 & 0.1520  &  0.4610(30) &  0.8141(124)  &  0.06219(136) & 150  & 7500 \\
  0.9000 & 0.1525  &  0.4167(52) &  0.7762(103) &  0.04920(99) & 150  & 7500 \\  
  0.9000 & 0.1530  &  0.3638(59) &  0.7145(166) & 0.03724(67) & 120  & 6000 \\ 
  \hline
  0.9500  & 0.1485 &0.4757(36)  &  0.8481(89) & 0.07469(57) & 170  & 8500  \\      
  0.9500  & 0.1490 & 0.4349(42) &  0.8207(75) & 0.06206(60) & 190  & 9500 \\    
  0.9500  & 0.1495 & 0.3858(46)  &  0.8069(63)  & 0.04809(115) & 160  & 8000 \\   
  0.9500  & 0.1500 & 0.3369(90) & 0.7210(212) & 0.03723(107) & 200  & 10000 \\ 
  \hline                                               
  1.000 & 0.1300   & 1.312(4)  &  0.9869(8)   &    0.5505(17)  & 30 & 3000 \\  
  1.000 & 0.1350  & 1.080(7)  &  0.9781(20)  &    0.3764(27)  & 30 & 3000 \\
  1.000 & 0.1400   & 0.8343(62)  &  0.9544(42)  &    0.2339(14)  & 30 & 3000 \\
  1.000 & 0.1450  & 0.5163(59)  &  0.8941(105) &    0.09890(239)  & 30 & 3000 \\
  1.000 & 0.1460  & 0.4444(51)  &  0.8596(78)  &    0.07472(137)  & 60 & 3000 \\
  1.000 & 0.1465  &  0.4108(58)& 0.8274(105)   &   0.06159(94)   & 170 & 8500  \\
  1.000 & 0.1470  & 0.3690(62)  &  0.7889(109) &    0.04959(158) & 120  & 6000 \\
  1.000 & 0.1475 & 0.3524(68)  &  0.7324(166) &  0.03512(154)  & 120   & 6000 \\ 
  \hline \hline
\end{tabular}
\caption{Summary of pseudoscalar (PS), vector (V),
 and partially conserved axial current (PCAC) masses
 measured on the lattice $16^3 \times 32$ for $\beta = 0.85, 0.90, 0.95, 1.00$.
The total numbers of configurations and of trajectories are also shown.}
 \label{app:masses2}
\end{table}

\acknowledgments

We are grateful to K.~Nagata and K.~Ishiguro for useful comments.
We also acknowledge the help of Cybermedia Center (CMC), Osaka University in tuning the gradient-flow code.
The numerical simulations were carried out on SX-ACE and OCTOPUS
 at CMC and Research Center for Nuclear Physics (RCNP), Osaka University,
 together with XC40 at Yukawa Institute for Theoretical Physics (YITP)
 and Institute for Information Management and Communication (IIMC), Kyoto University.
This work partially used computational resources of
 HPCI-JHPCN System Research Project (Project ID: jh200031) in Japan.
 This work was supported in part by Grants-in-Aid for Scientific Research 
through Grant Nos.
 15H05855,%(Nitta-san)
 18H01211, 18H05406, % IIda-san
and 19K03875%(Itou)
, which were provided by the Japan Society for the Promotion of Science (JSPS), and in part 
by the Program for the Strategic Research Foundation at Private Universities 
``Topological Science'' through Grant No.\ S1511006, which was supported by the
Ministry of Education, Culture, Sports, Science and Technology (MEXT) of Japan.
The work of E. I. is supported by the Iwanami-Fujukai Foundation.
T.-G. L. acknowledges the support of the Seiwa Memorial Foundation.

%\paragraph{Note added.} This is also a good position for notes added
%after the paper has been written.


\end{document}